\documentclass[%
 prl,
superscriptaddress,
twocolumn,
 amsmath,amssymb,
 aps,
]{revtex4-2}

\usepackage{graphicx}% Include figure files
\usepackage{dcolumn}% Align table columns on decimal point
\usepackage{bm}% bold math
\usepackage[colorlinks,linkcolor={blue},citecolor={blue},urlcolor={blue}]{hyperref}

\usepackage{eqnarray,amsmath}
\newcommand{\nhat}{\hat{\bf n}}
\newcommand{\uhat}{\hat{\bf u}}
\newcommand{\vhat}{\hat{\bf v}}

\newcommand{\Hhat}{\hat{\bf H}}

\newcommand{\xhat}{\hat{\bf x}}
\newcommand{\yhat}{\hat{\bf y}}

\newcommand{\nb}{\mathbf{n}}

\newcommand{\rb}{{\bf r}}
\newcommand{\ub}{{\bf u}}

\newcommand{\Eb}{{\bf E}}

\newcommand{\Hb}{{\bf H}}
\newcommand{\Ib}{{\bf I}}

\newcommand{\vb}{{\bf v}}

\newcommand{\Jb}{{\bf J}}

\newcommand{\omegab}{\mbox{\boldmath $\omega$\unboldmath}}

\newcommand{\nablab}{\mbox{\boldmath $\nabla$\unboldmath}} 

\newcommand{\Gammab}{\mbox{\boldmath $\Gamma$\unboldmath}}

\newcommand{\beq}{\begin{equation}}

\newcommand{\eeq}{\end{equation}}

\newcommand{\bea}{\begin{eqnarray}}

\newcommand{\eea}{\end{eqnarray}}

\usepackage[colorinlistoftodos]{todonotes}

\allowdisplaybreaks
\usepackage{orcidlink}

\begin{document}

%\preprint{APS/123-QED}

\title{Field-Particle Interactions in Curved Flows for Shape-Asymmetric Active Particles}

\author{Derek C. Gomes\orcidlink{0009-0006-6318-2624}}
\email{derekgomes@students.iisertirupati.ac.in}
\author{Tapan C. Adhyapak\orcidlink{0000-0002-8251-2880}}%
%\email{adhyapak@labs.iisertirupati.ac.in}
\email{Author to whom correspondence should be addressed: adhyapak@labs.iisertirupati.ac.in}
\affiliation{Department of Physics, Indian Institute of Science Education and
Research (IISER) Tirupati, Tirupati, Andhra Pradesh, 517619, India }


\begin{abstract}

We show that curvatures in general ambient flow profiles can align shape-asymmetric active particles, revealing a previously overlooked competition with externally applied aligning fields. Focusing on the ubiquitous
case of channel flows, we then investigate the fundamental consequences of this competition for the dynamics of shape-asymmetric active particles in
microchannels in the presence of orienting fields. We find that this interplay gives rise to novel mechanisms for controlling particle dynamics, with potentially broad applications, and suggests exciting possibilities such as active-particle analogs of electronic systems.

\end{abstract}

\maketitle

%\tableofcontents

Externally applied orienting fields induce rich microswimmer dynamics
\cite{Ybert2025MacroShear, Lauga2020BioconvectionMTB,
JabbariFarouji2025NoneqExtField, AdhikariPRL2020HydroMonopole,
ClementPRF2018MotorBrake, Koch2012ChemoInstability,
Brady2023ConfinedActMattFields, Peyla2020DeformableInGravity}, with promising
medical and biotechnological applications \cite{ Carvajal2024Reactor, Felfoul2016TumorDelivery, Clement2019RotorAssembly}. Consequently, controlling
microswimmers using orienting fields, often in combination with readily imposed
flows, has attracted considerable interest \cite{GoldsteinPRL2025, Ybert2016DestabilizationMagBac, salima_prl, BlakemoreScience1975}. Despite the demonstrated potential of field--flow
control, existing studies lack a unifying theoretical framework, limiting
predictive power and experimental design.

In this Letter, we formulate the leading-order flow-induced velocities of
head--tail shape-asymmetric (HT-A) particles in arbitrary flows and show that
local flow curvatures in distinct planes generate competing polar-alignment
torques. In the presence of activity, we investigate the interplay between
these torques and an external orienting field, uncovering a broad range of HT-A
microswimmer dynamics that can be tuned solely through the field strength and
orientation. These behaviors differ markedly from those of field-oriented
head--tail symmetric (HT-S) swimmers and field-free HT-A swimmers, opening new
avenues for microswimmer control and applications.

Many living microswimmers possess natural \cite{BlakemoreScience1975,
PedleyKesslerReview1992, LaugaBook2020} or bioengineered \cite{AubryACS2020}
dipole moments that couple to external fields. Field-coupled synthetic
microswimmers, enabling wide parameter control, are also gaining traction
\cite{ranabir_PRL2024,  Faivre2018Biohybrid, Bechinger2016Phototaxis,
Yu2019Thermophoresis}, with realizations based on magnetic, electric, and
gravitational fields, as well as light and chemical gradients.

Field-aligned microswimmers in flows exhibit striking phenomena, including
swimmer focusing and negative gravitaxis \cite{salima_prl, KesslerNature1985},
dense microbial layers \cite{DurhamKesslerStockerScience2009}, bioconvective
patterns \cite{Dervaux2022Phototaxis, Bees2020Bioconvection, ArrietaPRL2019}, and
microswimmer analogues of Bose-Einstein condensation
\cite{MengGolestanianPRL2021}. However, these studies are limited to HT-S
swimmers, leaving HT-A systems unexplored, where the flow-curvature-induced
alignment presented here may dominate.

\begin{figure}

\includegraphics[width=1.0\columnwidth]{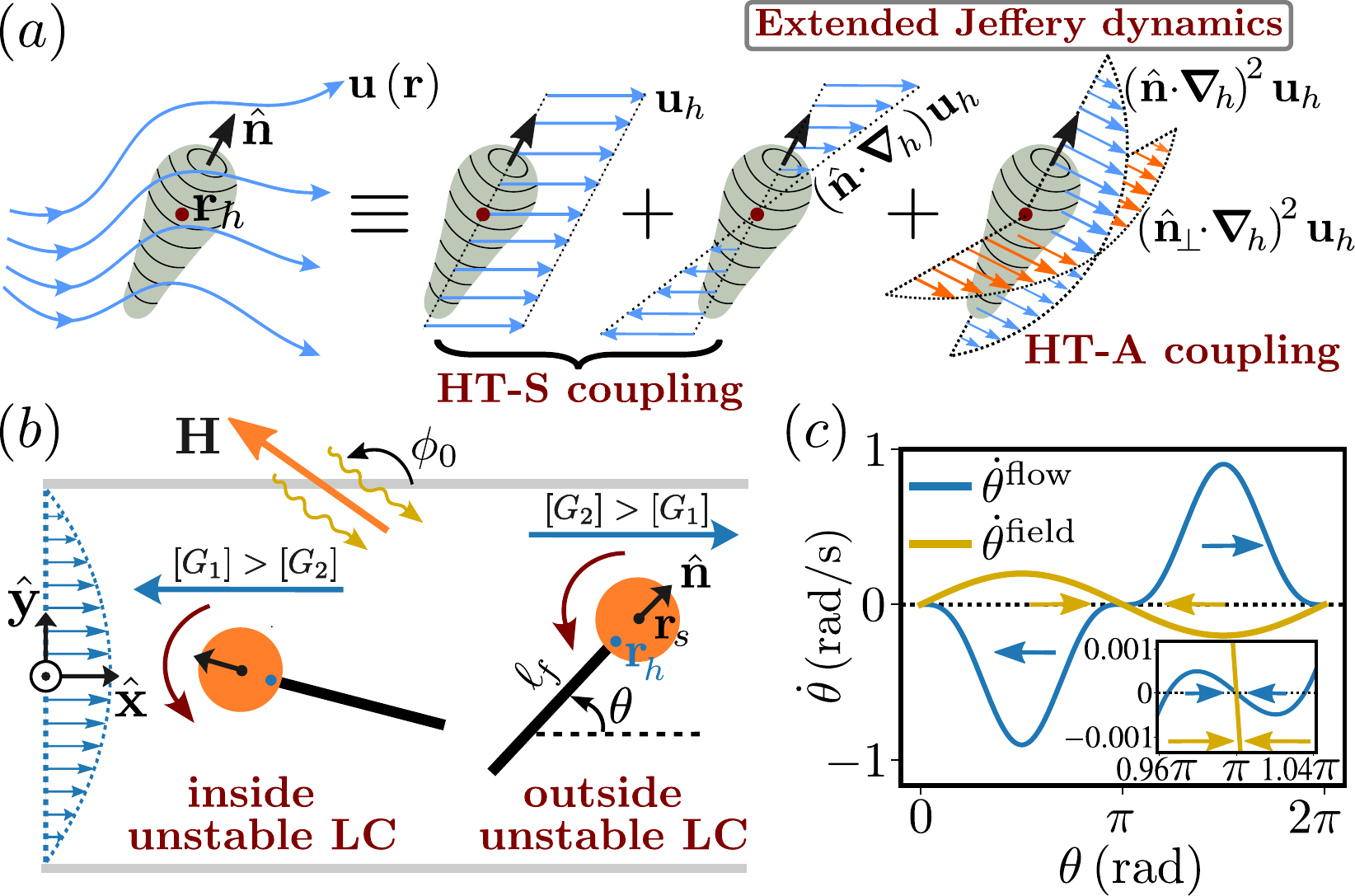}

\caption{(a) Coupling of an HT-A particle rotation to an arbitrary flow
$\ub(\rb)$. (b) Competing polar flow alignments ($G_1$ and $G_2$ terms for the
sphere-rod model) and their interplay with an external orienting field $\Hb$,
flow vorticity (maroon arrows), and apolar flow alignment (not shown).  (c)
Flow (blue) and field (yellow) contributions to $\dot{\theta}$ in Poiseuille
flow [see (b)] at $y=0$, $\phi_0=\pi$, and $\mathcal{H}=0.2\,\mathrm{s}^{-1}$.
Inset: magnified view near $\theta=\pi$.}

\label{Fig1}

\end{figure}

Most suspended active particles possess pronounced HT-A shapes. Prominent
examples include artificial swimmers \cite{Ambarish_magnetic_active,
Dreyfus_artificial_swimmer, Zhang2009ArtificialFlagella}, as well as living
microswimmers such as \emph{E.~coli} and \emph{B.~subtilis}, widely used as
prototypical active matter \cite{Aranson2022BacterialAM, Clement2026ScaleFree,
Wu2017WeakSync, Wioland2016VortexLattices, tapanPRE2015}. Our recent work
\cite{GomesAdhyapakPRL2025} showed that shape asymmetry can yield remarkable
dynamics. Nevertheless, most treatments of field-particle interactions in flows
rely on the Fax\'en--Jeffery equations \cite{PedleyKesslerReview1992,
Guasto2012Review, Faivre2024MagBacReview}, which implicitly assume HT-S
particles. Here, we generalize these equations to incorporate HT-A effects,
revealing the crucial role of shape asymmetry in field-coupled dynamics in
flows.

Our main findings are as follows. Oriented external fields focus HT-S
swimmers at arbitrary channel heights, unlike the midplane focusing or wall
escape of Ref.~\cite{salima_prl}. By exploiting polar flow alignment, HT-A
swimmers access focused, wall-going, oscillatory, and mixed states. A
field--flow competition explains the HT-A$\to$HT-S crossover, while a reduced
nonlinear theory reveals multiple field-controlled bifurcations that provide
additional control knobs.

\textit{Generalized Fax\'en--Jeffery Dynamics}---We consider an HT-A swimmer
suspended in an arbitrary imposed flow $\ub(\rb)$
[Fig.~\ref{Fig1}(a)]. At low Reynolds numbers, its hydrodynamic center $\rb_h$
and orientation $\nhat$ evolve as
$\dot{\rb}_h=v^{\rm sp}\nhat+\vb_h^{\rm flow}$ and
$\dot{\nhat}=\omegab_{h}^{\rm flow}\times\nhat$, where $v^{\rm sp}$ is the
self-propulsion speed in a quiescent fluid. Retaining all symmetry-allowed
terms to leading order in \emph{local} flow gradients, the flow-induced
velocities $\vb_h^{\rm flow}$ and $\omegab_h^{\rm flow}$ of a rigid HT-A
particle axisymmetric about $\nhat$ are
\begin{widetext}
\begin{eqnarray}    
&\vb^{\rm{flow}}_h   = \ub_h 
                        + \boldsymbol{\Gamma}_1\cdot \left(\nhat\cdot\nablab_h\right)\ub_h
                        + \boldsymbol{\Gamma}_2\cdot \nabla_h^2 \ub_h 
                        + \boldsymbol{\Gamma}_3 \cdot \left(\nhat\cdot\nablab_h\right)^2\ub_h, \label{eq:v_flow}\\
&\omegab_h^{\rm{flow}} = \frac{1}{2} \nablab_h \times \ub_h 
                          + \boldsymbol{\Gamma}_4\cdot\left(\nhat\cdot\nablab_h\right)\nablab_h\times \ub_h 
                          + \nhat \times 
                            \left[  \kappa_1 \Eb_h \cdot \nhat
                                  + \kappa_2 \nabla^2_h \ub_h
                                  + \kappa_3 \left( \nhat \cdot \nablab_h \right)^2 \ub_h\right].
\label{eq:omega_flow}
\end{eqnarray}
\end{widetext}
In the above, $\ub_h=\ub(\rb_h)$, $\nablab_h\equiv\nablab_{\rb_h}$, and
$\Eb_h=[\nablab_h\ub_h+(\nablab_h\ub_h)^T]/2$ is the strain-rate tensor at
$\rb_h$. The tensors
$\boldsymbol{\Gamma}_i=\Gamma_i^{\parallel}\nhat\nhat+\Gamma_i^\perp
(\Ib-\nhat\nhat)$ and the constants $\kappa_i$ depend on particle shape; their
values for the sphere-rod model used below are given in the Supplemental
Material (SM) \cite{supplemental_material}.

In Eq.~\eqref{eq:v_flow}, $\boldsymbol{\Gamma}_1$ is the only lowest-order HT-A
term that breaks $\nb\to-\nb$ symmetry; the others generalize the Fax\'en
theorem \cite{kim_karilla} to arbitrary axisymmetric shapes.  Likewise, in
Eq.~\eqref{eq:omega_flow}, $\boldsymbol{\Gamma}_4$, and $\kappa_{2,3}$ are the
only lowest-order HT-A terms that break fore-aft symmetry, while the rest
generalize the Jeffery equation \cite{JefferyPRSCA1922} to arbitrary
axisymmetric shapes.

For strong self-propulsion ($v^{\rm sp}\gg\|\vb_h^{\rm flow}\|$), HT-A
contributions to $\omegab_h^{\rm flow}$ dominate streamline crossing. In a
simplified picture [Fig.~\ref{Fig1}(a)], treating the $i$th segment centered at
$\rb_i$ on the symmetry axis as freely advected by
$\ub(\rb_i)+\epsilon_{\mu\nu}^i\nabla_\mu\nabla_\nu\ub(\rb_i)+\cdots$, and
expanding about $\rb_h$, shows that the first-order term
$(\nhat\!\cdot\!\nablab_h)\ub_h$ yields the classical Jeffery (HT-S) rotation
(vorticity and $\kappa_1$), while the leading HT-A couplings arise from local
flow curvatures, $\nabla_h^2\ub_h$ and $(\nhat\!\cdot\!\nablab_h)^2\ub_h$,
producing anisotropic rotations that distinguish curvatures in different planes
($\kappa_{2,3}$)~\cite{gamma4}.  Unlike the isotropic curvature couplings of
Ref.~\cite{Sriram_Active_Caustics}, our results rely on this anisotropic
mechanism.

\textit{Flagellated Swimmer in Imposed Field and Flow}---It is easy to see that
the $\kappa_{2,3}$ terms act as polar alignments, orienting the particle along
or against the local flow direction $\uhat$. External fields $\Hb$,
\textit{e.g.}, light or magnetic fields, coupled to the particle orientation
$\nhat$ compete with these terms, leading to rich field--flow dynamics.

To elucidate this interplay, we consider the minimal HT-A swimmer model of
Ref.~\cite{GomesAdhyapakPRL2025}: a spherical cell body of radius $a$ propelled
by a rigid, slender flagellar rod of length $\ell_f$, with an orientation
$\nhat$ directed from tail to head [Fig.~\ref{Fig1}(b)]. The model captures the
essential HT-A features of a broad class of biological
\cite{omori_upward_sperm_swim, Najafi2019MultipleBundles, Deng2020Pseudomonas,
tapanPRE2017} and synthetic \cite{Guan2023MagTadpole,
Shelley2019AuPtRods,Nelson2016Acoustic} flagellated microswimmers. We place the
swimmer in a Poiseuille flow, $\ub(\rb)=v_f[1-(\rb\cdot\yhat)^2/R^2]\xhat$,
corresponding to a microchannel of half-width $R$, ubiquitous in natural
\cite{Stocker2015MicrobesInFlow} and technological \cite{SquiresQuake2005,
StoneStroockAjdari2004} settings.

\begin{figure*}

\includegraphics[height=0.4\textwidth]{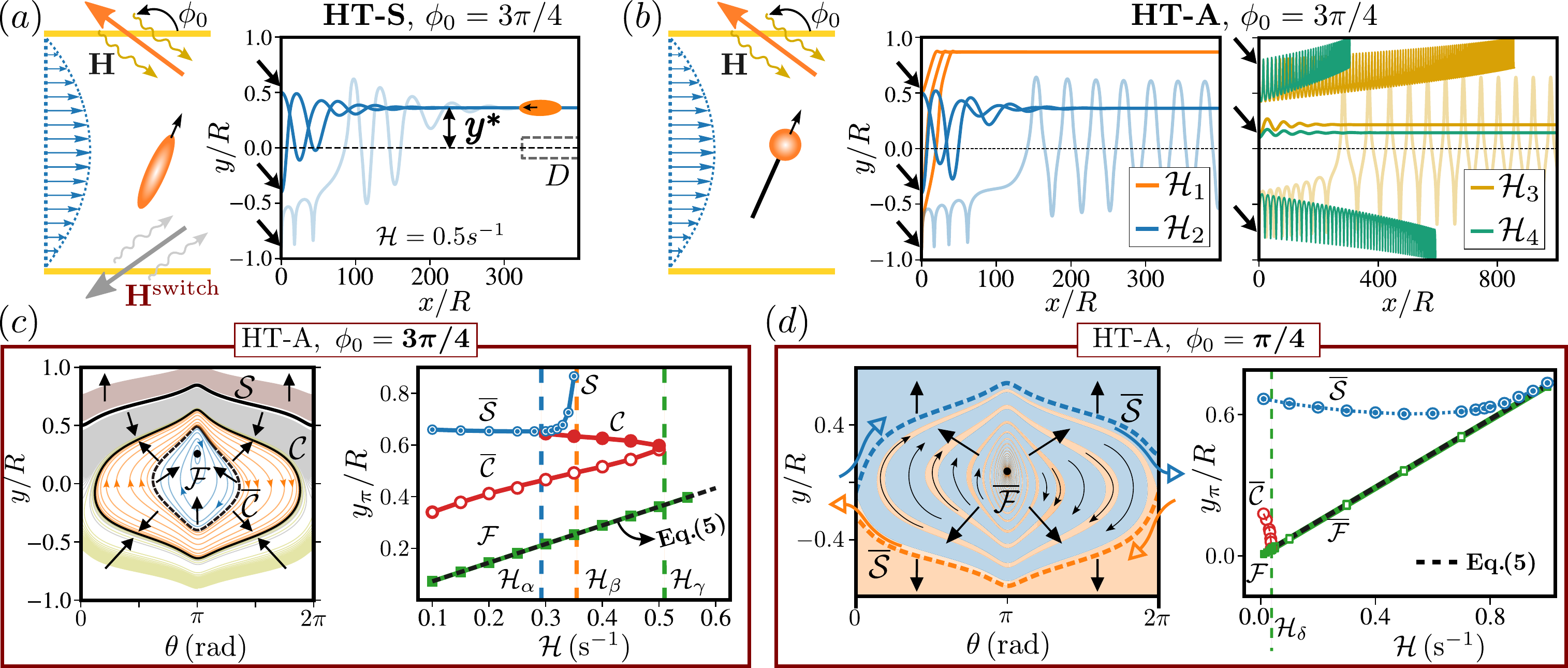}% Here is how to import EPS art

\caption{(a) Trajectories of an \textbf{HT-S swimmer} in Poiseuille flow for
different initial heights $y_i$ (colors) at $\theta_i=\pi$, under an applied
field $\Hb$ (schematic, left). \textbf{Living Transistor:} The swimmer current
through detector region $D$ is switched ON (OFF) by turning ON (OFF) an
auxiliary field $\Hb^{\rm switch}$. (b) Corresponding schematic (left) and
trajectories (middle, right) for an \textbf{HT-A swimmer} at decreasing field
strengths $\mathcal{H}_1>\cdots>\mathcal{H}_4$, with common initial heights
indicated by arrows; $\phi_0=3\pi/4$ and $\theta_i=\pi$. (c) \textit{Left:}
HT-A phase space for $\phi_0$ in the second quadrant at
$\mathcal{H}=0.35\,\mathrm{s}^{-1}\approx\mathcal{H}_3$, showing attractors
$\mathcal{F}$, $\overline{\mathcal{C}}$, $\mathcal{C}$, and separatrix
$\mathcal{S}$. \textit{Right:} Corresponding attractor heights $y_\pi$ versus
field strength, exhibiting sharp transitions at
$\mathcal{H}_{\alpha,\beta,\gamma}$. (d) Same as (c) for $\phi_0$ in the first
quadrant and $\mathcal{H}=0.1\,\mathrm{s}^{-1}$, revealing an additional
bifurcation at $\mathcal{H}_\delta$.}

\label{Fig2}
\end{figure*}

Experimentally, it is convenient to track the sphere center $\rb_s$ rather than
$\rb_h=\rb_s-r_{\rm sh}\nhat$, with $r_{\rm sh}>0$ [Fig.~\ref{Fig1}(b)]. The
dynamics becomes $\dot{\rb}_s=v^{\rm sp}\nhat+\vb_h^{\rm flow} +r_{\rm
sh}\omegab\times\nhat$ and $\dot{\nhat}=\omegab\times\nhat$, where
$\omegab=\omegab_h^{\rm flow}+\nhat\times\mathcal{H}\Hhat$ includes external
alignment. The scaled field $\mathcal{H}$ is related to the field strength $H$
by $H=\gamma_R\mathcal{H}/\mu$, where $\mu$ is a possible dipole moment and
$\gamma_R$ the rotational friction. In the high-P\'eclet-number regime,
assuming dominant shear-plane alignment \cite{GomesAdhyapakPRL2025,
shear-induced}, we set $\rb_s\equiv(x,y)$ and
$\nhat\equiv(\cos\theta,\sin\theta)$, with $\theta$ measured from $\uhat \equiv
\xhat$. Expanding $\ub_h$ and its gradients about $\rb_s$ \cite{supplemental_material} yields the
streamwise dynamics, $\dot{x}=[1+(a^2/6)\nabla^2_s]\ub(\rb_s)\cdot \xhat +
v^{\rm sp}\cos\theta-r_{\rm sh}\dot{\theta}\sin\theta + [\mathcal{A}_x -
\mathcal{A}_y \sin^2{\theta}]f(y,\theta)$, together with a coupled
$y$--$\theta$ dynamics independent of $x$:
\begin{eqnarray}
\dot{y}&=&     v^{\rm{sp}}\sin\theta + r_{\rm{sh}}\dot{\theta}\cos\theta
                + \mathcal{A}_y f (y,\theta)
                 \label{eq:ydot_poiseuille}\\
\dot{\theta} &=& \frac{v_f}{R^2} \left[ \left(1-G\right)  y
               + 2 G \sin^2{\theta} y + G_1 \sin{\theta}
               + G_2\sin^3{\theta}\right] \nonumber \\
                 && -\mathcal{H}\sin\left(\theta - \phi_0 \right).
                 \label{eq:thetadot_poiseuille}
\end{eqnarray}
Here, $\phi_0$ denotes the orientation of the applied field $\Hb$.  Expressions
for $f(y,\theta)$, and the anisotropy prefactors $\mathcal{A}_{x,y}$ are given
in SM~\cite{supplemental_material}.  The coefficient $G=\kappa_1$ is the
Jeffery constant, while $G_1=2\kappa_2$ and $G_2=2(\kappa_3-r_{\rm
sh}\kappa_1)$ characterize the polar curvature couplings.

Below, we use $v_f=500\,\mu\text{m}/\text{s}$ and $R=50\,\mu\text{m}$, with
$a=1\,\mu\text{m}$, $\ell_f=5\,\mu\text{m}$, $v^{\rm
sp}=50\,\mu\text{m}/\text{s}$, and $(\mathcal{A}_x,\mathcal{A}_y)=(0.538,
-0.127)$. For our swimmer, $G=0.902$ satisfies $0<G<1$, while 
$G_1=0.057\,\mu\text{m}>0$ and $G_2=-4.577\,\mu\text{m}<0$; accounting for active flows places $\rb_h$ at $r_{\rm
sh}=0.75a$ from $\rb_s$ \cite{GomesAdhyapakPRL2025}.  The equations are
integrated using Euler's method with $dt=5\times10^{-5}\,\text{s}$ to obtain
the trajectories $\{x(t),y(t),\theta(t)\}$.

\textit{Field vs. Flow Alignments}---Figure~\ref{Fig1}(c) illustrates the
competition between flow and field for an HT-A particle at $y=0$ with
$\phi_0=\pi$. The field-induced rotation, $\dot{\theta}^{\rm
field}=-\mathcal{H}\sin(\theta-\pi)$, always promotes upstream alignment
($\theta=\pi$), whereas the flow rotation $\dot{\theta}^{\rm
flow}=\omega_h^{\rm flow}$ contains competing $G_1$ and $G_2$ contributions
[see also Fig.~\ref{Fig1}(b)]. For most orientations, $G_2$ dominates,
destabilizing upstream alignment and opposing the field (main panel), while
near $\theta=\pi$, $G_1$ prevails, reinforcing upstream stabilization by the
field (inset).  This competition implies rich dynamics at low to intermediate
field strengths, while strong fields suppress polar flow alignment, driving
HT-A dynamics toward the HT-S limit ($G_1, G_2\,=0$). We therefore first consider
HT-S reference dynamics before turning to HT-A swimmers.

\emph{Field-Controlled Swimmer Focusing}---In the HT-S limit,
$\Gammab_1=\mathbf{0}$. Taking $\Gammab_{2,3}$ isotropic, with
$\Gammab_2=b^2\Ib/6$ and $\Gammab_3=(c^2-b^2)\Ib/6$, Eq.~\eqref{eq:v_flow}
reduces to Fax\'en's equation for an ellipsoid \cite{kim_karilla}, where
$c=b\sqrt{(1+G)/(1-G)}$ gives Jeffery coefficient $G$. We use
$b=1\,\mu\mathrm{m}$. Accordingly, Eq.~\eqref{eq:ydot_poiseuille} becomes
$\dot{y}=v^{\rm sp}\sin\theta$. The crucial rotational dynamics follows from
Eq.~\eqref{eq:thetadot_poiseuille} with $G_1=G_2=0$, yielding a field-augmented
Jeffery equation. Combined with translation, it leads to our central prediction
of field-tunable swimmer focusing.

Figure~\ref{Fig2}(a) shows HT-S trajectories for $\phi_0=3\pi/4$ in the second
quadrant, initialized at $(y_i,\,\theta_i = \pi)$. All focus to a finite height
$y^*\neq0$, unlike the midplane focusing of Refs.~\cite{salima_prl,
MengGolestanianPRL2021}, while maintaining upstream orientation. Notably,
focusing persists for arbitrary $(y_i,\theta_i)$, collapsing a dilute
population into a swimmer beam at $y^*$.

Similar behavior occurs throughout the second and third $\phi_0$-quadrants,
with swimmers focusing toward the field-side half of the channel, with $y^*$
increasing (decreasing) as $\phi_0\to\pi/2\,(3\pi/2)$. By contrast, in the
first and fourth quadrants, trajectories approach the walls, with a larger
fraction reaching the field-side wall (for uniformly distributed initial
conditions). This asymmetry vanishes as $\phi_0\to0$, recovering
Ref.~\cite{salima_prl}.

Remarkably, Eqs.~\eqref{eq:ydot_poiseuille} and
\eqref{eq:thetadot_poiseuille} admit the fixed point (FP)
\begin{eqnarray}
\left(y^*,\theta^*\right)=\left(\mathcal{H}\sin\phi_0\,\frac{R^2/v_f}{1-G},\;\pi\right),
\label{eq:FP}
\end{eqnarray}
which is stable in the second and third $\phi_0$ quadrants [see below
Eq.~\eqref{extended-VdP}]~\cite{saddle}. Besides explaining off-midplane
focusing, it predicts a focus height linear in $\mathcal{H}$ and tunable
through both the field strength and orientation---one of our key findings. In
the first and fourth quadrants, the FP is unstable, yielding wall-approaching
trajectories.

\textit{HT-A Particle Dynamics}---Figure~\ref{Fig2}(b) shows HT-A trajectories
at $\phi_0=3\pi/4$ (second quadrant), revealing dynamics far richer than for
HT-S swimmers. The focused state of Eq.~\eqref{eq:FP} persists at all
$\mathcal{H}$. At large fields ($\mathcal{H}_1=1.2\,\mathrm{s}^{-1}$), the
dynamics reduces to the single-state HT-S limit. Decreasing $\mathcal{H}$
successively produces mixed states: at
$\mathcal{H}_2=0.5\,\mathrm{s}^{-1}$, stable midplane-crossing oscillations
coexist with focusing; at $\mathcal{H}_3=0.3\,\mathrm{s}^{-1}$, an additional
state of oscillations approaching only the field-side wall emerges. At
$\mathcal{H}_4=0.2\,\mathrm{s}^{-1}$, the stable oscillations disappear and a
new far-side-wall state appears, with most trajectories reaching either wall
and a small fraction focusing near $(y^*\!\to\!0,\pi)$, as in the
$\mathcal{H}\!\to\!0$ limit of Ref.~\cite{GomesAdhyapakPRL2025}.

The phase space at $\mathcal{H}=0.35\,\mathrm{s}^{-1}$ [Fig.~\ref{Fig2}(c),
left], encompassing all initial conditions, contains a stable FP
$\mathcal{F}$ and stable LC $\mathcal{C}$, corresponding to the focused and
stable oscillatory states of Fig.~\ref{Fig2}(b). The separatrix $\mathcal{S}$
bounds wall-approaching trajectories, while the unstable LC
$\overline{\mathcal{C}}$ separates focused and oscillatory dynamics.

The evolution of these structures with $\mathcal{H}$ is tracked by their height
$y_\pi$ at $\theta=\pi$ [Fig.~\ref{Fig2}(c), right]. For
$\mathcal{H}>\mathcal{H}_\gamma$, only the FP $\mathcal{F}$ survives,
recovering the HT-S limit. At $\mathcal{H}_\gamma$, a sharp
\textit{\textbf{bifurcation}} creates the stable--unstable LC pair
$(\mathcal{C},\overline{\mathcal{C}})$, which separate as $\mathcal{H}$
decreases until $\mathcal{H}_\alpha$, where $\mathcal{C}$ collides with the
separatrix $\mathcal{S}$ emerging at $\mathcal{H}_\beta$
($\mathcal{H}_\alpha<\mathcal{H}_\beta<\mathcal{H}_\gamma$). For
$\mathcal{H}<\mathcal{H}_\alpha$, only $\mathcal{F}$, $\overline{\mathcal{C}}$,
and $\overline{\mathcal{S}}$ remain, recovering the
$\mathcal{H}\!\to\!0$ dynamics of Ref.~\cite{GomesAdhyapakPRL2025}.

The dynamics is analogous in the second and third $\phi_0$ quadrants. By
contrast, in the first and fourth quadrants [Fig.~\ref{Fig2}(d), left], no
focused or stable oscillatory states exist: the FP $\overline{\mathcal{F}}$ is
unstable, no stable LC appears, and all trajectories approach the walls, with
blue (orange) regions reaching the field- (far-) side wall. The separatrix
$\overline{\mathcal{S}}$ separates trajectories with and without midplane
crossing, while the orange region shrinks with increasing $\mathcal{H}$. The
phase-space topology [Fig.~\ref{Fig2}(d), right] remains otherwise unchanged,
except below $\mathcal{H}_\delta$, where $\overline{\mathcal{F}}$ stabilizes
($\overline{\mathcal{F}}\!\to\!\mathcal{F}$) and an unstable LC
$\overline{\mathcal{C}}$ emerges, marking a \textit{\textbf{subcritical Hopf
bifurcation}} to the $\mathcal{H}\!\to\!0$ limit.

\begin{figure}

\includegraphics[width=\columnwidth]{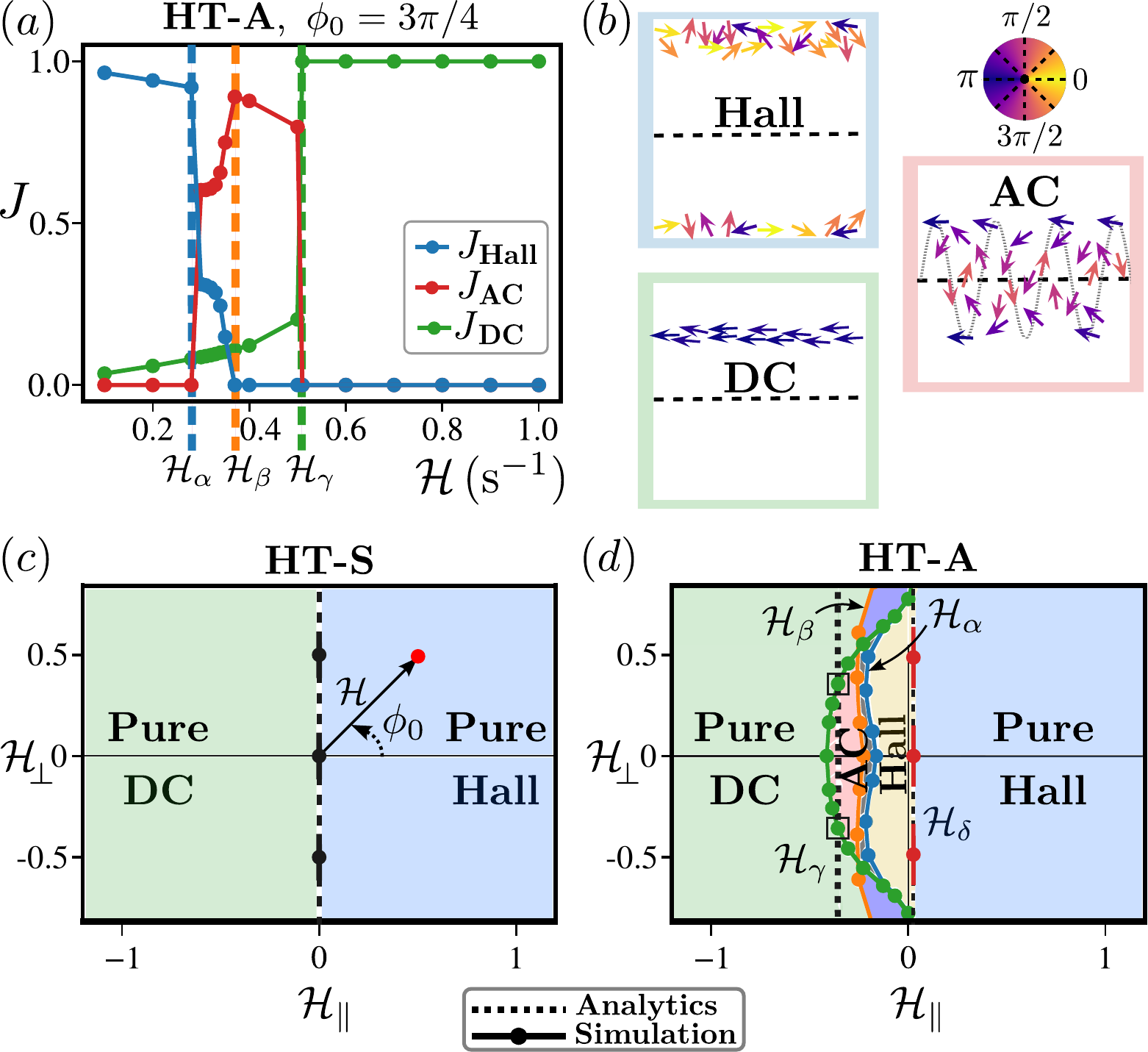}

\caption{\textit{\textbf{Current characteristics and dynamical phases}}: (a)
$J$ versus $\mathcal{H}$ for HT-A swimmers at $\phi_0=3\pi/4$, showing Hall,
DC, and AC current regimes. The transition fields
$\mathcal{H}_{\alpha,\beta,\gamma}$ coincide with Fig.~\ref{Fig2}(c). (b)
Schematic swimmer populations in the Hall, DC, and AC regimes; arrow colors
indicate swimmer orientations. (c),(d) Phase diagrams in the
$\mathcal{H}_{\parallel}$--$\mathcal{H}_{\perp}$ plane [$(\mathcal{H},\phi_0)$
in polar coordinates] for HT-S and HT-A swimmers, respectively, where
$\mathcal{H}_{\parallel}=\mathcal{H}\cos\phi_0$ and
$\mathcal{H}_{\perp}=\mathcal{H}\sin\phi_0$.  Pure phases occur for all
$(y_i,\theta_i)$; otherwise, the phase labels denote the dominant state reached
by the majority of initial conditions.  Dashed and solid curves denote
analytical and simulation phase boundaries, respectively.}

\label{Fig3}
\end{figure}

\textit{$J$--$\mathcal{H}$ Characteristics and Phase Diagrams}---To
characterize the field response of a dilute suspension, we classify
trajectories as stable oscillatory (AC), focused (DC), or wall-approaching
(Hall, with mean drift transverse to the flow). The corresponding
velocity-normalized partial currents are $\Jb_\mu=J_\mu\vhat_\mu$ (no sum over
$\mu$), where $J_\mu$ is the fractional population and $\vhat_\mu$ the mean
drift direction for $\mu\in\{\mathrm{AC,DC,Hall}\}$.

We estimate $J_\mu$ from the phase-space fraction of each class in the
$y$--$\theta$ plane [Fig.~\ref{Fig2}(c),(d)]. For HT-S swimmers, the response
is purely DC in the second $\phi_0$-quadrant and purely Hall in the first, for
all $\mathcal{H}$ (see Fig.~S1 in the SM~\cite{supplemental_material}). By
contrast, HT-A swimmers exhibit Hall-, AC-, and DC-dominated regimes within the
same quadrant ($\phi_0=3\pi/4$) [Fig.~\ref{Fig3}(a)], separated by
$\mathcal{H}_{\alpha,\gamma}$, with the Hall current vanishing at
$\mathcal{H}_\beta$; these coincide with the transitions in Fig.~\ref{Fig2}(c).
The corresponding populations are shown in Fig.~\ref{Fig3}(b).

Figures~\ref{Fig3}(c),(d) summarize the full $\mathcal{H}$--$\phi_0$ phase
behavior for HT-S and HT-A swimmers, respectively. Notably, the HT-A topology
contains a mixed region at small $\mathcal{H}_{\parallel} =
\mathcal{H}\cos\phi_0$, where multiple states coexist, with the labeled state
dominating in each sub-region.

We now rationalize these observations analytically.

\textit{Stability and Control of Dynamical Attractors}---Eliminating $y$
from Eqs.~\eqref{eq:ydot_poiseuille} and \eqref{eq:thetadot_poiseuille} maps
the coupled $y$--$\theta$ dynamics onto the $\theta$--$\dot{\theta}$ phase
space, yielding a single second-order equation for the orientational dynamics,
$\ddot{\theta}=\ddot{\theta}(\theta,\dot{\theta})$.  Linearizing this equation
about the FP $(y^*,\theta^*)\equiv(\theta^*,\dot{\theta}^*)$ of
Eq.~\eqref{eq:FP}, with $\psi\equiv\theta-\theta^*$ and
$\dot{\psi}\equiv\dot{\theta}-\dot{\theta}^*$, gives
\begin{eqnarray}
&&\ddot{\psi}+k\psi=\zeta_0\dot{\psi},
\label{extended-VdP}
\end{eqnarray}
where $\zeta_0=\mathcal{H}\cos\phi_0- \left[G_1+(1-G)r_{\rm sh}\right]v_f/R^2$
and $k>0$~\cite{supplemental_material}.

Equation~\eqref{extended-VdP} admits the FP $(\psi^*,\dot{\psi}^*)=(0,0)$,
corresponding to $(y^*,\theta^*)$ in the $y$--$\theta$ plane. Its stability
is set by $\zeta_0$: for $\zeta_0<0$, the equation reduces to a
damped harmonic oscillator and the FP is stable; for $\zeta_0>0$, it becomes
pumped and the FP unstable. These results apply to both HT-S and HT-A
swimmers.

For HT-S particles ($G_1=r_{\rm sh}=0$), $\zeta_0=\mathcal{H}\cos\phi_0$.
Hence, the FP is stable only in the second and third quadrants and unstable
otherwise, accounting for the focusing (DC) and wall-going (Hall) regimes in
Fig.~\ref{Fig3}(c). Remarkably, even with full nonlinearities, HT-S dynamics
admits no additional attractors [Fig.~\ref{Fig2_Appendix}], implying that its
phase diagram is entirely determined by the linear theory.

For HT-A swimmers, nonlinearities generate additional attractors (see below),
while the linear theory still captures the stable FP $\mathcal{F}$ in
Fig.~\ref{Fig2}(c) and unstable FP $\overline{\mathcal{F}}$ in
Fig.~\ref{Fig2}(d). Indeed, $\zeta_0=\mathcal{H}\cos\phi_0-[G_1+(1-G)r_{\rm
sh}]v_f/R^2$ is negative in the second and third quadrants but may become
positive in the first and fourth, with $\zeta_0=0$ giving the
$\mathcal{F}\!\to\!\overline{\mathcal{F}}$ transition at
$\mathcal{H}_\delta=v_f[G_1+(1-G)r_{\rm sh}]/(R^2\cos\phi_0)$. For
$\phi_0=\pi/4$, this yields $\mathcal{H}_\delta=0.037\,\mathrm{s}^{-1}$, in
excellent agreement with Fig.~\ref{Fig2}(d).

\textit{Nonlinear Effects and HT-A Phase Diagram}---Following End Matter,
Eq.~\eqref{extended-VdP}, including all nonlinear terms
[Eq.~\eqref{eq:nlin_oscillator}], reduces to the \emph{bifurcation normal form}
$\dot{\psi}_a=-r\psi_a+u\psi_a^3-v\psi_a^5$, where $\psi_a$ is the $\psi$
amplitude. Its nonzero root magnitudes, $|\psi_a^{\pm}| =
\left[(u\pm\sqrt{u^2-4rv})/(2v)\right]^{1/2}$, correspond to the stable and
unstable limit cycles ($\mathcal{C}$, $\overline{\mathcal{C}}$) in
Fig.~\ref{Fig2}(c), respectively. These annihilate at $u^2=4rv$
(\textit{\textbf{saddle-node bifurcation}}), yielding an analytical estimate of
$\mathcal{H}_{\gamma}$, beyond which HT-A swimmers abruptly enter the HT-S
regime with purely DC currents [Fig.~\ref{Fig3}(d)]. The transition fields
$\mathcal{H}_{\alpha,\beta}$ and further details of the phase diagram are given
in End Matter.

\textit{Conclusion}---Our theory of HT-A swimmer dynamics captures the
interplay of imposed flows and external fields beyond classical
Fax\'en--Jeffery dynamics, predicting field-tunable focusing heights for both
HT-S and HT-A swimmers. Most remarkably, while HT-S swimmers exhibit only
focused populations, HT-A swimmers can be tuned between focused and extended
oscillatory populations by varying $(\mathcal{H},\phi_0)$. A reduced nonlinear
theory further explains the phase behavior, the field-driven HT-A$\to$HT-S
crossover, and the associated transition fields.

The framework extends directly to non-Poiseuille flows with varying
curvature, \textit{e.g.}, peristaltic and cilia-driven flows, where HT-S and
HT-A swimmers are predicted to focus at distinct heights. Together with the
phase diagrams, these results suggest shape-based swimmer sorting and
active-matter devices based on field-controlled transport, including the
\emph{living-transistor} concept of Fig.~\ref{Fig2}(a).

%===================================================

\emph{Acknowledgments}---TCA and DCG thank Sriram Ramaswamy for insightful
comments and discussions.  TCA acknowledges grants CRG/2021/004759 from the
ANRF (India) and MoE-STARS/STARS-2/2023-0814 for financial support. DCG
acknowledges PMRF, Govt.  of India, for fellowship and funding. TCA and DCG
express gratitude to the \emph{Indian Institute of Science Education and
Research (IISER) Tirupati} for funds and facilities.  Furthermore, the support
and the resources provided by `PARAM Brahma Facility' under the National
Supercomputing Mission, Government of India at the \emph{Indian Institute of
Science Education and Research (IISER) Pune} are gratefully acknowledged.

%\bibliography{bibDerek}

\begin{thebibliography}{61}
\bibitem [{gam()}]{gamma4}%
  \BibitemOpen
  \href@noop {} {}\bibinfo {note} {The HT-A contribution associated with
  $\boldsymbol{\Gamma}_4$ originates from vorticity gradients but, for our
  model, is absorbed into the vorticity evaluated at $\rb_s$ that is displaced
  from $\rb_h$.}\BibitemShut {Stop}%
\bibitem [{sad()}]{saddle}%
  \BibitemOpen
  \href@noop {} {}\bibinfo {note} {The equations also admit a saddle point
  $(-y^*,0)$ for both HT-S and HT-A swimmers, for $\phi_0$ in all quadrants. It
  captures an infinitesimal population at $-y^*$ and forms boundary points of a
  separatrix $\overline{\mathcal{S}}$, similar to that in Fig.~\ref{Fig2}(d),
  which underlie the wall-population asymmetry for first- and fourth-quadrant
  $\phi_0$.}\BibitemShut {Stop}%
\end{thebibliography}

%============================================================
% End: main.bbl
%============================================================

\providecommand{\noopsort}[1]{}\providecommand{\singleletter}[1]{#1}%
%

%============================================================
% End: main.bbl
%============================================================

\newpage

\appendix
\renewcommand{\theequation}{A\arabic{equation}}
\setcounter{equation}{0}

\renewcommand{\thefigure}{A\arabic{figure}}
\setcounter{figure}{0}

\onecolumngrid

\subsection{End Matter}

\twocolumngrid

\textit{Nonlinear Attractors: Theory and Derivations}---

We derive the bifurcation normal form presented in the main text, from which
the LCs and the approximate HT-A$\to$HT-S crossover are obtained.
Here, we further estimate the remaining transition fields, analyze the
separatrix, and establish a quantitative mechanism governing the crossover.
Including all nonlinearities, Eq.~\eqref{extended-VdP} of the main text becomes
\begin{eqnarray}
\ddot{\psi} +  k\psi + f_{\rm{anh}}\left( \psi\right)
                 &=&  \zeta_0 \dot{\psi} + \zeta_L\left( \psi \right) \dot{\psi}
                    + \zeta_Q \left( \psi \right) \dot{\psi}^2 \nonumber \\
                 &\equiv& \zeta(\psi,\dot{\psi})\dot{\psi},
\label{eq:nlin_oscillator}
\end{eqnarray}
a nonlinear oscillator with anharmonic function
$f_{\rm anh}(\psi)$ and nonlinear damping/pumping functions
$\zeta_{L,Q}(\psi)$ listed in the
SM~\cite{supplemental_material}. Unlike
Ref.~\onlinecite{GomesAdhyapakPRL2025}, the present equation admits
field-induced non-convergent solutions and therefore requires a substantially
different treatment.

\begin{figure}

\includegraphics[width=\columnwidth]{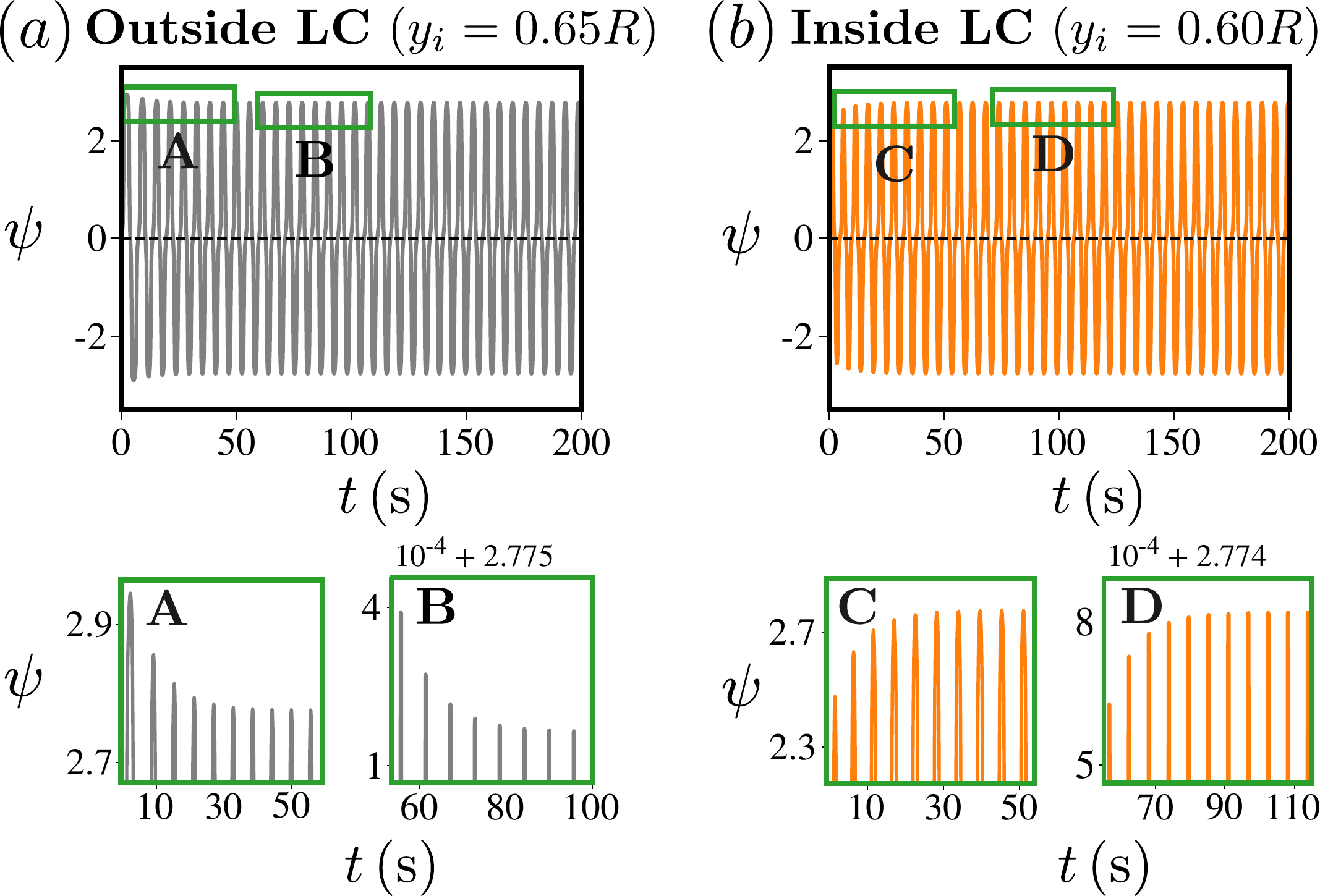}

\caption{(a) Time evolution of $\psi=\theta-\pi$ for a simulated $y$--$\theta$ trajectory
approaching the stable LC $\mathcal{C}$ from outside [gray curve in
Fig.~\ref{Fig2}(c)]. (b) Same as (a), but for a trajectory approaching
$\mathcal{C}$ from inside [orange curve in Fig.~\ref{Fig2}(c)]. Bottom panels:
enlarged views of time windows A--B [(a)] and C--D [(b)] (green boxes),
illustrating the slow approach to $\mathcal{C}$. In both cases, $\theta_i=\pi$
($\psi=0$).}

\label{Fig1_Appendix}
\end{figure}

We adopt the standard two-time ansatz \cite{strogatz2018nonlinear},
\begin{eqnarray}
\psi=\psi_a(\tau)\cos\left[\sqrt{k}t+\varphi(\tau)\right],
\label{eq:psi_ansatz}
\end{eqnarray}
with $\tau\ll\sqrt{k}\,t$. As illustrated in Fig.~\ref{Fig1_Appendix}, the
solution captures the fast oscillations and slowly varying amplitude $\psi_a$
of the simulated $\psi$ dynamics. As $y$--$\theta$ trajectories approach the LC
$\mathcal{C}$, $\psi_a$ approaches a FP.  More generally, nonzero FPs of
$\psi_a$ determine the existence of LCs, their nature determines LC stability,
and their values set the LC sizes, whereas the FP at $\psi_a=0$ corresponds to
the focused state $(y^*,\pi)$.  Thus, the FP and LC attractors are completely
characterized by the $\psi_a$ dynamics.

\textit{$\dot{\psi}_a$ vs. $\psi_a$ equation}---Substituting the ansatz
[Eq.~\eqref{eq:psi_ansatz}] into Eq.~\eqref{eq:nlin_oscillator}, neglecting
second derivatives of the slow variables $\psi_a$ and $\varphi$, and
rearranging, we obtain
\begin{eqnarray}
\dot{\psi}_a \sin A + \dot{\varphi}\,\psi_a \cos A
                    &=& \frac{1}{2\sqrt{k}}f_{\rm anh}
                        + \frac{1}{2}\zeta\,\psi_a \sin A,~~~~~
\label{eq:nlin_psia_1}
\end{eqnarray}
where $A=\sqrt{k}t+\varphi$, and we approximate $\dot{\psi} \approx -
\sqrt{k}\,\psi_a\sin A$; $f_{\rm anh}$ and $\zeta$ are functions of $\psi_a\cos
A$ and $\psi_a\sin A$. Multiplying by $\sin A$ and averaging over one period
($A:0\to2\pi$) eliminates the $\dot{\varphi}$ term.  Since the Taylor expansion
of $f_{\rm anh}$ contains only powers of $\cos A$, $\langle f_{\rm anh}\sin
A\rangle=\langle\cos^pA\,\sin A\rangle=0$ for all integers $p$.
Equation~\eqref{eq:nlin_psia_1} therefore reduces to
\begin{eqnarray}
\dot{\psi}_a
= \left\langle\zeta\,\sin^2A\right\rangle \psi_a.
\label{eq:nlin_psia_2}
\end{eqnarray}

\emph{Pad\'{e} Approximation:} Taylor expansion of $\zeta$ in
Eq.~\eqref{eq:nlin_psia_2} yields terms of the form
$\langle\cos^p\!A\,\sin^q\!A\rangle$, which are generally nonzero, giving an
infinite series in $\psi_a$ on the RHS. However, the series is nonconvergent
and cannot be truncated. We therefore approximate it by a
\emph{Pad\'{e} approximant}, i.e., the ratio $P_m/Q_n$ of polynomials $P_m$ and
$Q_n$ of degrees $m$ and $n$, respectively, with the constant term of $Q_n$
fixed to unity~\cite{Baker1996Pade}. Equation~\eqref{eq:nlin_psia_2} then
becomes
\begin{eqnarray}
\dot{\psi}_a \approx
             \frac{P_m(\psi_a)}{Q_n(\psi_a)}
             =\frac{p_0+p_1\psi_a+\cdots+p_m\psi_a^m}
                    {1+q_1\psi_a+\cdots+q_n\psi_a^n},
\label{eq:nlin_psia_3}
\end{eqnarray}
where the coefficients $p_i(\Hb,v_f/R^2)$ and $q_i(\Hb,v_f/R^2)$ depend on the
field and flow, and are determined by matching the first $m+n$ terms of the
infinite series following the standard Pad\'e construction. We take $m=9$ and
$n=3$, which, among the lowest polynomial degrees, best reproduces the
numerically observed transition field $\mathcal{H}_\gamma$ at $\phi_0=3\pi/4$,
where the stable and unstable nonzero FPs of Eq.~\eqref{eq:nlin_psia_3} merge
[Fig.~\ref{Fig2}(c) and black squares in Fig.~\ref{Fig3}(d)].

\begin{figure}
\includegraphics[width=\columnwidth]{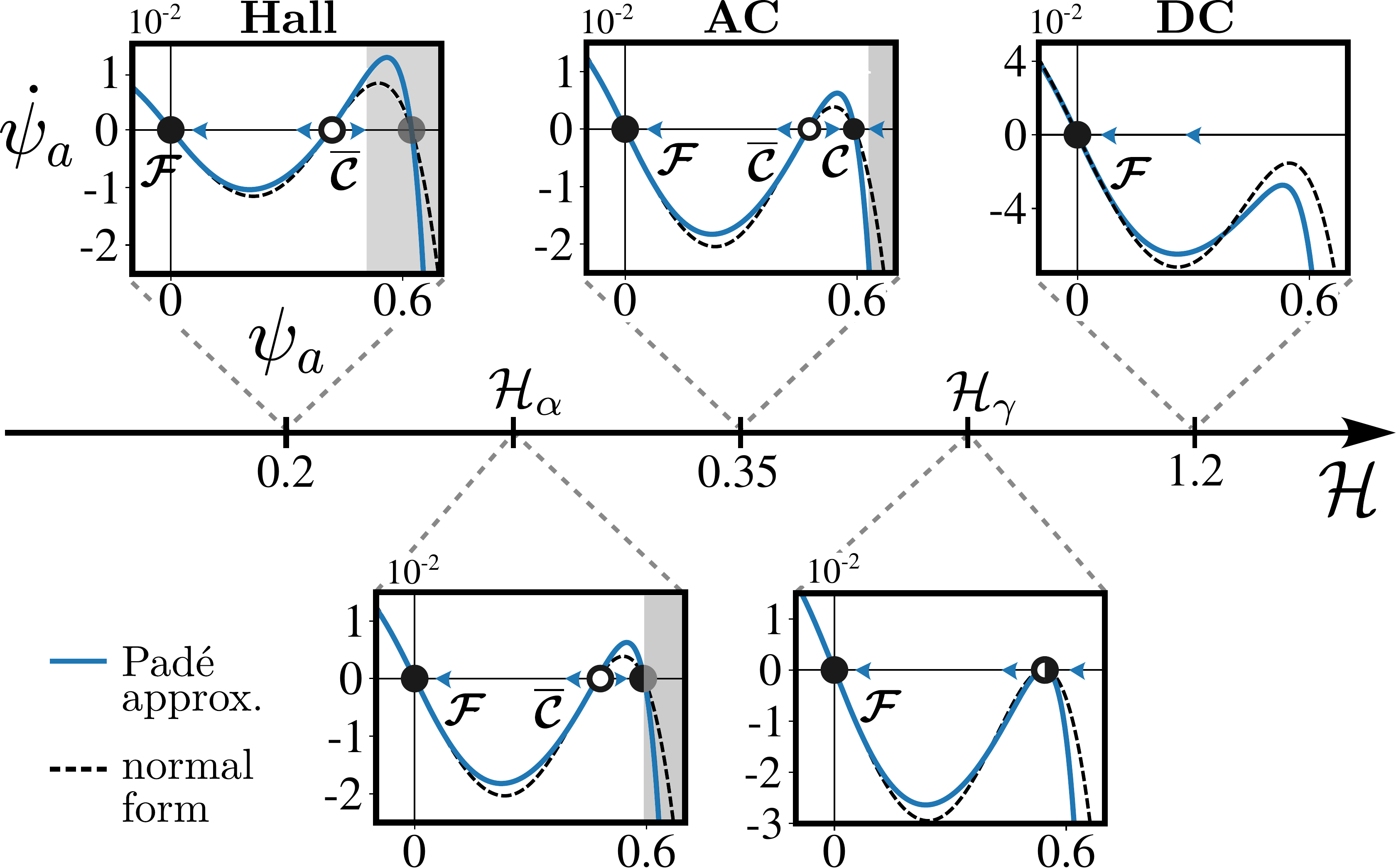}

\caption{(Top) Analytical $\dot{\psi}_a$--$\psi_a$ curves for representative
$\mathcal{H}$ values in the Hall, AC, and DC regimes. Arrows and circles denote
FP stability (filled: stable; open: unstable; half-filled: half-stable). The
shaded region beyond $\psi_a^{\rm max}$ [Eq.~\eqref{eq:psia_max}] is
inaccessible to physical LCs owing to the separatrix. The DC panel illustrates
the HT-S dynamics reached by HT-A swimmers at large $\mathcal{H}$. (Bottom)
Same as above, but at the transition fields $\mathcal{H}_{\alpha,\gamma}$. In
all panels, $\phi_0=3\pi/4$; the $\mathcal{H}$ axis is not to scale.}

\label{Fig2_Appendix}
\end{figure}

\textbf{\emph{Transition Fields and Emergence of LCs:}} Equation
\eqref{eq:nlin_psia_3} yields three distinct FPs, $\mathcal{F}$,
$\overline{\mathcal{C}}$, and $\mathcal{C}$. For $\phi_0=3\pi/4$, these are
shown in Fig.~\ref{Fig2_Appendix}, corresponding to the FP and the unstable and
stable LCs of Fig.~\ref{Fig2}(c), respectively.  Remarkably, the variation of
the FP separations with $\mathcal{H}$ qualitatively captures the numerical
trend observed in the full dynamics [right panel, Fig.~\ref{Fig2}(c)]. In
particular, the LC FPs, $\overline{\mathcal{C}}$ and $\mathcal{C}$, approach,
merge, and annihilate at $\mathcal{H}_\gamma$, identifying the AC$\to$DC
transition as a \textit{\textbf{saddle-node bifurcation}} [the
$\mathcal{H}_\gamma$ and DC panels, Fig.~\ref{Fig2_Appendix}]. Evaluating
Eq.~\eqref{eq:nlin_psia_3} for different $\phi_0$, and identifying the field at
which $\overline{\mathcal{C}}$ and $\mathcal{C}$ merge, yields the theoretical
$\mathcal{H}_\gamma(\phi_0)$ boundary in Fig.~\ref{Fig3}(d), providing a good
estimate of the Pure DC regime.

\textbf{\emph{Bifurcation normal form:}} Equation~\eqref{eq:nlin_psia_3} can be
further approximated by the fifth-order polynomial
\begin{eqnarray}
\dot{\psi}_a \approx -r\psi_a+u\psi_a^3-v\psi_a^5,
\label{eq:nlin_psia_4}
\end{eqnarray}
the simplest approximation that preserves the physical FP structure. Here,
$r,u,v>0$ are chosen to match the slope at the origin and the root locations of
$P_m/Q_n$. Equation~\eqref{eq:nlin_psia_4} is the
\emph{bifurcation normal form} presented in the main text. The resulting
$\dot{\psi}_a$--$\psi_a$ curves agree well with those of
Eq.~\eqref{eq:nlin_psia_3} [Fig.~\ref{Fig2_Appendix}]. The three FPs occur at
$\psi_a=0$, corresponding to $\mathcal{F}$, and
\[
\psi_a^{\mp}
=\left[\frac{u\mp\sqrt{u^2-4rv}}{2v}\right]^{1/2},
\]
corresponding to the unstable and stable LCs
($\overline{\mathcal{C}},\mathcal{C}$), respectively. As noted in the main
text, the \textit{\textbf{saddle-node bifurcation}} is then given by the
compact relation
\begin{eqnarray}
u^2 = 4rv,
\label{eq:fold_bifurcation}
\end{eqnarray}
for which $\psi_a^{+}=\psi_a^{-}$. The analytical field and flow dependences of
$r$, $u$, and $v$ are, in principle, obtainable from those of
$p_i(\Hb,v_f/R^2)$ and $q_i(\Hb,v_f/R^2)$.  Finally, it is important to note
that Eq.~\eqref{eq:nlin_psia_4} cannot be obtained by a direct fifth-order
polynomial approximation of Eq.~\eqref{eq:nlin_psia_2}, since the latter is
nonconvergent.

\textbf{\emph{Separatrix and Physical LCs:}} Besides the analytically
determined LCs, separatrices also exist, but arise from strong nonlinearities
and are therefore obtained only from simulations. Their height,
$y_\pi^{\mathcal{S}}$, estimates the maximum allowed LC height since LCs cannot
cross a separatrix [see, {\it e.g.}, Fig.~\ref{Fig2}(c), main text]. The
corresponding LC width, $\psi_a^{\rm max}$, therefore bounds the physically
accessible nonzero FPs of $\psi_a$. We estimate $\psi_a^{\rm max}$ from the LC
equation $\psi=\psi^{\rm LC}(y)$, obtained by substituting
Eq.~\eqref{eq:psi_ansatz} with constant $\psi_a,\varphi$ into
Eq.~\eqref{eq:thetadot_poiseuille}$,$ and evaluating the result at
$y=y_\pi^{\mathcal{S}}$:
\begin{eqnarray}
\psi_a^{\rm max}
=\frac{1}{\sqrt{k}}
\left[\frac{v_f}{R^2}(1-G)y_\pi^{\mathcal S}
-\mathcal{H}\sin\phi_0\right].
\label{eq:psia_max}
\end{eqnarray}
This bound defines the inaccessible LC region shaded in
Fig.~\ref{Fig2_Appendix} and explains the Hall$\to$AC transition at
$\mathcal{H}_\alpha$, where $\psi_a^{+}=\psi_a^{\rm max}$ [$\mathcal{H}_\alpha$
panel, Fig.~\ref{Fig2_Appendix}]. For $\psi_a^{+}<\psi_a^{\rm max}$, the stable
LC $\mathcal{C}$ becomes accessible, giving rise to AC currents
[Fig.\,\ref{Fig3}(a)].

The remaining transition field, $\mathcal{H}_{\beta}$, occurs when the
separatrix height exceeds the channel radius [Fig.~\ref{Fig2}(c)]. Since this
transition involves only the separatrix, which arises from strong
nonlinearities, $\mathcal{H}_{\beta}$ is determined solely from simulations.

\end{document}